\documentclass[11pt,a4paper]{article}
\pdfoutput=1

\usepackage{amsmath}
\usepackage{amsfonts}
\usepackage{amsthm}
\usepackage{amssymb}
\usepackage{amscd}
\usepackage[british]{babel}
\usepackage{graphicx}
\usepackage{psfrag}
\usepackage{epsfig}
\usepackage{rotating}
\usepackage{times}
\usepackage{color}
\usepackage{ulem}

\theoremstyle{plain}

\newtheorem{remark}{Remark}[section]

\newcommand{\boxend}{\flushright{$\Box$}}

\begin{document}

\title{Quasi-matter domination parameters in bouncing cosmologies}

\author{Emilio Elizalde$^{a,}$\footnote{E-mail: elizalde@ieec.uab.es},  Jaume Haro$^{b,}$\footnote{E-mail: jaime.haro@upc.edu} and  Sergei D. Odintsov$^{a,c,}$
\footnote{E-mail: odintsov@ieec.uab.es}
}

\maketitle

\begin{center}
{\small $^{a}$Instituto de Ciencias del Espacio (CSIC) and
Institut d'Estudis Espacials de Catalunya (IEEC/CSIC), Campus Universitat
Aut\`{o}noma de Barcelona \\ Torre C5-Parell-2a planta, 08193 Bellaterra
(Barcelona) Spain\\

$^b$Departament de Matem\`atica Aplicada I, Universitat
Polit\`ecnica de Catalunya \\ Diagonal 647, 08028 Barcelona, Spain \\

$^{c}$
%Instituto de Ciencias del Espacio (CSIC) and
%Institut d'Estudis Espacials de Catalunya (IEEC/CSIC), Campus Universitat
%Aut\`{o}noma de Barcelona \\ Torre C5-Parell-2a planta, 08193 Bellaterra
%(Barcelona), Spain\\
Instituci\'o Catalana de Recerca i Estudis Avan\c cats (ICREA), Barcelona,
Spain
}
\end{center}

\thispagestyle{empty}

\begin{abstract}

For bouncing cosmologies, a fine set of parameters is introduced in order to describe the nearly matter dominated phase, and which play the same role 
that the usual slow-roll parameters play in inflationary cosmology. It is shown that, as in the inflation case, the spectral index and the running parameter 
for scalar perturbations in bouncing cosmologies can be best expressed in terms of these small parameters. Further, they explicitly exhibit the duality which 
exists between a nearly matter dominated Universe in its contracting phase and the quasi de Sitter regime in the expanding one. The results obtained also confirm 
and extend the known evidence that the spectral index 
for an exactly matter dominated Universe (i.e., a pressureless Universe) in the contracting phase
is, in fact, the same as the spectral index 
for an 
exact de Sitter regime in the expanding phase. Finally, in both the inflationary and the matter bounce scenarios, the theoretical values 
of the spectral index and of the running parameter are compared with
their experimental counterparts, obtained from the most recent PLANCK data, with the result that the bouncing models here discussed do fit well accurate astronomical observations.

\end{abstract}

\vspace{0.5cm}

{\bf Pacs numbers:} 98.80.-k, 98.80.Jk, 05.50.Kd

%\vspace{0.5cm}

\section{Introduction}

Matter bounce scenarios \cite{brandenberger} are characterized by the Universe being matter dominated at very early times in the contracting phase and  evolving towards a bounce, to 
enter into an expanding regime, where it matches the behavior of the standard hot Friedmann Universe. They constitute a viable alternative to the inflationary paradigm.

 It is also well-known that matter domination in the contracting phase leads to the same spectral index, $n_s$, as for the case of the de Sitter regime in the expanding 
 Universe, namely $n_s=1$ \cite{wands}.
 This value does not agree with   the experimental one,   $n_s=0.9603\pm 0.0073$, which has been obtained from the most recent PLANCK data \cite{Ade}. In contrast, this 
 observational value can actually be accounted for in inflationary cosmology,
 because the Universe does not inflate following an exactly de Sitter regime. Instead, the inflaton field slow-roll in its potential drives the Universe to a quasi
 de Sitter stage. In such slow-roll regime, the leading perturbative term of the spectral index depends on two small parameters, so-called {\it slow-roll parameters}
 \cite{Stewart}, which are obtained explicitly as functions of the potential
 and its derivatives. By conveniently fitting these parameters one is able to match the theoretical value of the spectral index with the corresponding experimental one.

 Following the inflationary paradigm, in order to obtain a correct theoretical value of the spectral index in a matter bounce scenario---when we consider a single scalar 
 field only---we will introduce at very early times in the contracting phase some  dimensionless parameters, which we will  call  {\it quasi-matter domination parameters}.
 When these parameters are less than 1, the Universe will be nearly matter dominated in the contracting phase, in exact analogy with the inflationary Universe
case, where a small value of the  slow-roll parameters leads to  a  Universe in the expanding phase,  near the de Sitter regime.

 The aim of the present work is to construct viable bouncing cosmologies where the matter part of the Lagrangian is composed of a scalar field and, therefore, have to go 
 beyond  General Relativity,
 since the flat Friedmann-Lema{\^\i}tre-Robertson-Walker (FLRW) geometry  forbids  bounces when one deals
 with a single field (recall that bounces are allowed for FLRW geometries with a positive spatial curvature \cite{Clesse}).
  Hence for the flat FLRW geometry, theories such as  holonomy corrected Loop Quantum Cosmology \cite{wilson}, where a big bounce appears 
 owing to the discrete structure of space-time \cite{ashtekar}, teleparalellism \cite{Haro1}, or modified $F(R)$ gravity \cite{Haro2} must be taken into account.
  When dealing with these theories, in order to obtain a theoretical value of the
 spectral index that may fit well with current experimental data, a quasi-matter dominated regime in the contracting phase has 
 to be introduced, which is conveniently fixed by the quasi-matter domination parameters.
Moreover, in slow roll inflation one also considers the running of the spectral index
corresponding
to $N$ e-folds before the end of the inflation, which
in general, is  of the order of $N^{-2}$.
This
value turns out to be very small, when one substitutes for $N$ the minimum number of e-folds which are needed to solve the horizon and flatness problem in inflationary 
cosmology ($N> 50$), as compared with its corresponding observational value
$-0.0134\pm 0.009$ coming from the most recent PLANCK data \cite{Ade}, what shows that these slow roll models are less favored by observations. In contrast, in
matter bounce scenarios the number of e-folds
before the end of the quasi-matter domination regime
can be relatively small, for the horizon problem does not exist in bouncing cosmologies and the flatness problem is neutral {}{\cite{brandenberger1}}. This gives ground for the 
viability of such models, making thus possible that for certain matter bounce scenarios the theoretical values of the spectral index and of the running parameter do agree 
well with PLANCK observations.

 \section{Quasi-matter domination parameters}

 In General Relativity, for a flat Friedmann-Lema\^{\i}tre-Robertson-Walker (FLRW) geometry the Friedmann and conservation equations for a single  scalar field are
 \begin{eqnarray}\label{x1}
  H^2=\frac{1}{3}\left(\frac{\dot{\varphi}^2}{2}+V\right);\quad \ddot{\varphi}+3H\dot{\varphi}+V_{\varphi}=0.
 \end{eqnarray}
Assuming, in the contracting phase, quasi-matter domination at early times, i.e., $\dot{\varphi}^2\cong 2V\Longrightarrow\ddot{\varphi}\cong V_{\varphi}$, these  equations become
 \begin{eqnarray}\label{x2}
  \left\{\begin{array}{ccc}
          H^2&=& \frac{2}{3}V\\
          3H\dot{\varphi}+2V_{\varphi}&=&0
         \end{array}
\right.\Longleftrightarrow
\left\{\begin{array}{ccc}
         {\mathcal H}^2&=& \frac{2}{3}a^2V\\
          3{\mathcal H}{\varphi}'+2a^2V_{\varphi}&=&0.
         \end{array}
\right.
 \end{eqnarray}

Now, in complete analogy to the slow-roll regime in inflationary cosmology, we define our {\it  quasi-matter domination parameters} as
\begin{eqnarray}\label{x3}
 \bar\epsilon={}{
 -1-\frac{2}{3}\frac{\dot{H}}{H^2}= -\frac{2}{3}\left( \frac{1}{2}+\frac{{\mathcal H}'}{{\mathcal H}^2} \right)
 \cong \frac{1}{3}\left(\frac{V_{\varphi}}{V} \right)^2-1},
\end{eqnarray}
%\begin{eqnarray}\label{x5}
% \bar\eta= 2\bar{\epsilon}-\frac{\dot{\bar\epsilon}}{2H\left(\frac{3}{2}+\bar\epsilon\right)}\cong3\left(\frac{V_{\varphi\varphi}}{3V}-1\right),
%\end{eqnarray}
\begin{eqnarray}
 \bar{\delta}^2={}{
 %2\bar{\epsilon}-{\bar\eta}
 \frac{\dot{\bar\epsilon}}{2H\left(1+\bar\epsilon\right)}}
 \cong  -\left(\frac{V_{\varphi}}{V} \right)_{\varphi},
\end{eqnarray}
and
\begin{eqnarray}
\bar{\xi}^3=
 -\frac{1}{H}{\frac{d{\bar{\delta}}^2}{dt}} \cong-\frac{V_{\varphi}}{V}
\left(\frac{V_{\varphi}}{V} \right)_{\varphi\varphi},
\end{eqnarray}
which characterize this regime through the condition that $|\bar\epsilon|\ll 1$.
%(Note that, since $\bar\eta=2\bar\epsilon-\bar\delta^2$, the condition $|\bar\epsilon|\ll 1$
%implies $|\bar\eta|\ll 1$).
%\begin{eqnarray}\label{x4}
% \quad \bar\delta=-\frac{3}{2}-\frac{\ddot{H}}{2H\dot{H}} =-\frac{3}{2}-\frac{\ddot{\varphi}}{H\dot{\varphi}}=-\frac{1}{2}-\frac{{\varphi}''}{{\mathcal H}{\varphi}'}=
% \bar\eta-\bar\epsilon.\end{eqnarray}

In view of subsequent calculations, it is important to obtain the evolution of the parameters $\bar\epsilon$ and $\bar{\delta}^2$, which is given by
\begin{eqnarray}\label{derivades}
  \dot{\bar\epsilon}\cong {}{2}H\bar{\delta}^2,\quad   \frac{d{\bar{\delta}}^2}{dt} = -H\bar{\xi}^3.
\end{eqnarray}

 Since a potential of the form $e^{-\sqrt{{3}}|\varphi|}$ generates exact matter-domination, we will re-express our potential $V$, for negative values of the field,
 as  $V(\varphi)=e^{\sqrt{{3}}\varphi}W(\varphi)$, thus obtaining
 \begin{eqnarray}\label{z1}
  \bar\epsilon=
  %\cong\sqrt{3}\frac{W_{\varphi}}{W}+\frac{1}{2}\left(\frac{W_{\varphi}}{W}\right)^2
 {}{ \frac{2}{\sqrt{3}}}\frac{W_{\varphi}}{W},\quad
  %\bar\eta\cong2\sqrt{3}\frac{W_{\varphi}}{W}+\frac{W_{\varphi\varphi}}{W}\cong 2\sqrt{3}\frac{W_{\varphi}}{W},\nonumber\\
  \bar\delta^2\cong -\left(\frac{W_{\varphi}}{W}\right)_{\varphi},\quad
  \bar\xi^3 \cong
  %-6\left(\frac{W_{\varphi}}{W}\right)_{\varphi}
  -\sqrt{3} \left(\frac{W_{\varphi}}{W}\right)_{\varphi\varphi}   ,
 \end{eqnarray}
what means that, for a very flat potential $W$, these parameters are very small and nearly constant. In fact, since the expressions in (\ref{z1}) resemble those of the
slow-roll parameters, we conclude that we can choose $W$ as the new potential, namely the same potential as is used in slow-roll inflation. Note also that,
from (\ref{z1}) one gets {}{the following hierarchy}  $|\bar\xi^3|\ll |\bar\delta^2|\ll  |\bar\epsilon|$.

 As an example, for the potential $W(\varphi)=\lambda\varphi^{2n}$ one has
 \begin{eqnarray}
  \bar\epsilon={}{\frac{4n}{\sqrt{3}\varphi}+\frac{n^2}{3\varphi^2}\cong \frac{4n}{\sqrt{3}\varphi}},
  %\quad\bar\eta=\frac{4n\sqrt{3}}{\varphi}+\frac{n(n-1)}{\varphi^2}\cong \frac{4n\sqrt{3}}{\varphi}\nonumber\\
  \quad \bar\delta^2=-\frac{2n}{\varphi^2},\quad \bar\xi^3=
  %-\frac{6n}{\varphi^2}+
  \frac{4\sqrt{3}n}{\varphi^3}.
  %\cong -\frac{6n}{\varphi^2}.
 \end{eqnarray}
One can also introduce, in the same way as in the inflation setup, the number of e-folds before the end of the quasi-matter domination period, as follows $a(N)=e^{N}a_f$,
where $a_f$ is the value of the scale factor at the end of this regime.

With this definition, in the quasi-matter approximation the number of e-folds can be calculated as
\begin{eqnarray}
 N=-\int_{t_f}^{t_N}H(t)dt\cong \int_{\varphi(N)}^{\varphi_f}\frac{V}{V_{\varphi}}d\varphi,
\end{eqnarray}
which, in terms of the potential $W$, becomes
\begin{eqnarray}
 N\cong \int_{\varphi(N)}^{\varphi_f}\frac{1}{\sqrt{3}+\frac{W_{\varphi}}{W}}d\varphi.
\end{eqnarray}
For the particular case of the potential  $W(\varphi)=\lambda\varphi^{2n}$, for instance, one has
 \begin{eqnarray}
  N\cong -\frac{1}{\sqrt{3}}(\varphi(N)-\varphi_f)-\frac{2n}{3}\ln\left|\frac{\varphi_f+\frac{2n}{\sqrt{3}}}{\varphi(N)+\frac{2n}{\sqrt{3}}} \right|.
 \end{eqnarray}

Choosing the value of $\varphi_f$ when $\bar\epsilon=-1$, one finally obtains:
\begin{eqnarray}
  N\cong -\frac{1}{\sqrt{3}}\left(\varphi(N)+\frac{4n}{\sqrt{3}}\right)-\frac{2n}{3}\ln\left|\frac{4n}{\sqrt{3}\varphi(N)+{2n}} \right|.
 \end{eqnarray}

\subsection{The spectral index in bouncing cosmologies}

It is well-known that when one considers a scalar field only, General Relativity dealing with the flat FLRW geometry forbids bounces from the contracting
to the expanding phase, what is best seen by looking to
the Raychaudury equation $\dot{H}=-\frac{1}{2}\dot{\varphi}^2<0$: as the Hubble parameter always decreases, it is absolutely
impossible to pass from negative to positive values. For this reason,
when the matter part of the Lagrangian is given in terms of a single scalar field, one is led to use cosmologies beyond the
realm of General Relativity as, e.g.,  Loop Quantum Cosmology, teleparallel $F(T)$ gravity,
or $F(R)$ gravities.

Common to all these cases is
the Mukhanov-Sasaki \cite{sasaki} equation for scalar perturbations, in Fourier space, which can be expressed as
\begin{eqnarray}\label{MS}
 v_k''+\left(k^2-\frac{z''}{z}\right)v_k=0,
\end{eqnarray}
where, for very low energy densities and curvatures, $z=a\frac{\dot{\varphi}}{H}=a\frac{{\varphi}'}{{\mathcal H}}$. The explicit expressions for $z$ in the cases of $F(T)$ and 
$F(R)$ gravities have been obtained, respectively, in \cite{haro} \cite{Hwang}.

To derive the expression of $\frac{z''}{z}$ in the contracting phase, during the quasi-matter domination happening at very low energy densities and curvatures,  first we calculate
\begin{eqnarray}\label{x6}
 \frac{z'}{{\mathcal H}z}=
 %1+2\bar\epsilon-\bar\eta=
 1+\bar\delta^2.
\end{eqnarray}

Now, using the same method as in \cite{riotto} (pgs. 54-55), and the second formula of (\ref{derivades}), we obtain
\begin{eqnarray}\label{x7}
\frac{z''}{z}= {\mathcal H}(\bar\delta^2)'+{\mathcal H}' \frac{z'}{{\mathcal H}z}+{\mathcal H}^2\left(\frac{z'}{{\mathcal H}z}\right)^2
\nonumber\\
\cong-{\mathcal H}^2\bar\xi^3+
{\mathcal H}' (1+\bar\delta^2)+{\mathcal H}^2(1+2\bar\delta^2).
\end{eqnarray}
Finally,
solving the equation (\ref{x3}) for $\bar\epsilon$ constant (because $\frac{d\bar\epsilon}{dN}=\frac{\dot{\bar\epsilon}}{H}\cong {}{2}\bar\delta^2\ll \bar\epsilon$), i.e.,
taking ${\mathcal H}=\frac{2}{\eta(1+{}{3}\bar\epsilon)}\cong \frac{2}{\eta}(1-{}{3}\bar\epsilon)$
and replacing this expression in (\ref{x7}), we get, up to first order,
\begin{eqnarray}\label{x7}
\frac{z''}{z}
%\cong \frac{2}{\eta^2}(1-6\bar\epsilon-\bar\xi^2-4\bar\delta^2)
\cong \frac{2}{\eta^2}(1-{}{9}\bar\epsilon).
\end{eqnarray}

It is clear from this result that the Mukhanov-Sasaki
equation (\ref{MS}), during the quasi-matter domination epoch, can be approximated by
\begin{eqnarray}\label{x10}
  v_k''+\left(k^2-  \frac{1}{\eta^2}\left(\nu^2-\frac{1}{4} \right)\right)v_k=0\quad\mbox{ where }\quad \nu\cong \frac{3}{2}-{}{6}\bar\epsilon.
\end{eqnarray}
Then, in order to obtain the adiabatic Bunch-Davies vacuum,
one has to choose as a solution of (\ref{x10})
\begin{eqnarray}\label{hankel}
 v_k=\frac{\sqrt{\pi|\eta|}}{2}e^{i(1+2\nu)\frac{\pi}{4}}H^{(1)}_{\nu}(k|\eta|).
\end{eqnarray}
For modes well outside of the Hubble radius $k|\eta|\ll 1$,  (\ref{x10}) becomes
\begin{eqnarray}\label{x12}
  v_k''- \frac{1}{\eta^2}\left(\nu^2-\frac{1}{4} \right)v_k=0,
\end{eqnarray}
which solution is given by
\begin{eqnarray}\label{y13}
 v_k=C_1(k)|\eta|^{\frac{1}{2}+\nu}+C_2(k)|\eta|^{\frac{1}{2}-\nu}\cong C_2(k)|\eta|^{\frac{1}{2}-\nu}.
\end{eqnarray}

On the other hand, if one chooses as a scale factor in the quasi-matter domination period $a(t)\cong t^{2/3}\Longrightarrow a\cong \frac{\eta^2}{9}
\Longrightarrow z\cong \frac{\eta^2}{3\sqrt{3}}$, the solution (\ref{y13}) can be written as follows
\begin{eqnarray}\label{zz1}
 v_k\cong
 %\bar{C}_1(k) z(\eta)|\eta|^{-\frac{3}{2}+\nu}+
 \frac{1}{\sqrt{3}}{C}_2(k)\left(z(\eta)\int_{-\infty}^{\eta} \frac{d\bar\eta}{z^2(\bar\eta)}\right)|\eta|^{\frac{3}{2}-\nu}.
\end{eqnarray}

For modes well outside of the Hubble radius the solution (\ref{hankel}) should match (\ref{zz1}). Using the small argument approximation in the Hankel function
and the expression (\ref{y13}), we get for these modes
\begin{eqnarray}
 v_k\cong -i\sqrt{\frac{1}{6}}k^{-3/2}e^{i(1+2\nu)\frac{\pi}{4}}
 \frac{\Gamma(\nu)}{\Gamma(3/2)}\left( z(\eta)\int_{-\infty}^{\eta} \frac{d\bar\eta}{z^2(\bar\eta)} \right)\left(\frac{k|\eta|}{2} \right)^{\frac{3}{2}-\nu}.
\end{eqnarray}
Such modes will re-enter the Hubble radius at late times in the expanding phase, when the Universe is matter dominated. Then, the power spectrum is given by
\begin{eqnarray}\label{aa1}
 {\mathcal P}_S(k)=\frac{1}{12\pi^2}\left(\frac{\Gamma(\nu)}{\Gamma(3/2)}\right)^2\left(\int_{-\infty}^{\eta} \frac{d\bar\eta}{z^2(\bar\eta)}\right)^2\left(\frac{k}{aH} \right)^{3-2\nu},
\end{eqnarray}
where we have used the matter-domination condition, i.e., the relation $aH=\frac{2}{\eta}$.

Evaluating this quantity at the re-entry time ($aH=k$) and taking into account that this happens at very late times, we obtain the final formula for the power spectrum corresponding to 
scalar perturbations:
\begin{eqnarray}
 {\mathcal P}_S(k)=\frac{1}{12\pi^2}\left(\int_{-\infty}^{+\infty} \frac{d\eta}{z^2(\eta)}\right)^2_{k=aH},
\end{eqnarray}
where  the approximation ${\Gamma(\nu)}\cong {\Gamma(3/2)}$ has been performed.

\begin{remark}
Note that in slow-roll inflation, the power spectrum could be expressed in terms of the slow roll parameter $\bar{\epsilon}_{sr}$. This is due to the fact that
  in inflationary cosmology, for modes well outside of the Hubble radius,  the dominant mode  is constant being  
the other one  decreasing in the expanding phase. For this reason, one could write the power spectrum in terms of $\bar{\epsilon}_{sr}$, because it only depends  on the slow roll regime. 
Unfortunately, when one deals with 
bouncing cosmologies, in the contracting phase, for modes well outside of the Hubble radius, the dominant mode is not the constant one, because the other one 
increases. Then, the spectrum depends on the whole background evolution and not only on the quasi-matter domination regime. 
\end{remark}

In our case, the spectral index for scalar perturbations, namely $n_s$, is obtained from (\ref{aa1}) giving, as a result,
\begin{eqnarray}
 n_s-1\equiv\frac{\ln {\mathcal P}(k)}{\ln k}=3-2\nu={}{12}\bar\epsilon.
\end{eqnarray}

We can also calculate the running of the spectral tilt
\begin{eqnarray}
 \alpha_s\equiv \left(\frac{dn_s}{d\ln k}\right)_{k=aH}=\frac{n_s'}{(\ln aH)'}=-\frac{2n_s'}{{\mathcal H}}\cong-\frac{{}{24}\bar\epsilon'}{{\mathcal H}}=-48\bar\delta^2,
\end{eqnarray}
where we have used the formula (\ref{x3})  and the first formula of (\ref{derivades}).

In terms of the pressure and energy density, $P$ and $\rho$, respectively, one has
\begin{eqnarray}
\bar\epsilon=\frac{P}{\rho},\quad \bar\delta^2=\frac{1}{2H}\frac{d}{dt}\ln\left(1+\frac{P}{\rho} \right)
\end{eqnarray}
which leads to the
equivalent expression for the spectral index and the running parameter
%and in terms, of the energy density and pressure, is given by
\begin{eqnarray}\label{espectral_index}
 n_s-1=   12\frac{P}{\rho},\quad \alpha_s=-\frac{24}{H}\frac{d}{dt}\ln\left(1+\frac{P}{\rho} \right).
 %-\frac{3}{2}\frac{d}{Hdt}\ln\left(1+\frac{P}{\rho} \right)
 %-{3}\frac{d}{Hdt}\left(\frac{P}{\rho}\right)+\frac{d}{Hdt}\left(\frac{d}{2Hdt}\ln\left(1+\frac{P}{\rho} \right)\right)\right\}.
\end{eqnarray}

In the same way, for tensor perturbations one obtains the following power spectrum
\begin{eqnarray}
 {\mathcal P}_T(k)=\frac{2}{9\pi^2}\left(\int_{-\infty}^{+\infty} \frac{d\eta}{z_T^2(\eta)}\right)^2_{k=aH},
\end{eqnarray}
where, for very low energy densities and curvatures, $z_T=a$.
The exact expression of $z_T$ in holonomy corrected Loop Quantum Cosmology was obtained in \cite{Cailleteau}, in teleparallel $F(T)$ gravity in \cite{haro},
and in modified $F(R)$ gravity in \cite{Hwang1}, respectively.

The ratio of tensor to scalar perturbations is given by
\begin{eqnarray}\label{ratio}
 r=\frac{8}{3}\left(\frac{\int_{-\infty}^{+\infty} \frac{d\eta}{z_T^2(\eta)}}{\int_{-\infty}^{+\infty} \frac{d\eta}{z^2(\eta)}}   \right)^2_{k=aH}.
\end{eqnarray}

Finally,
it is  instructive to compare these parameters with the slow-roll ones commonly used in inflation:
\begin{eqnarray}\label{x14}
\bar\epsilon_{sr}= -\frac{\dot{H}}{H^2}\cong \frac{1}{2}\left(\frac{V_{\varphi}}{V} \right)^2, \quad \bar\eta_{sr}=2\bar\epsilon_{sr}-\frac{\dot{\bar\epsilon}_{sr}}{2H\bar\epsilon_{sr}}
\cong\frac{V_{\varphi\varphi}}{V},
\end{eqnarray}
which are related with the quasi-matter domination parameters $\bar{\epsilon}$ and $\bar{\delta}^2$ via the formulas
\begin{eqnarray}
 \bar\epsilon_{sr}=\frac{3}{2}(\bar{\epsilon}+1),\quad \bar\eta_{sr}=3(\bar{\epsilon}+1)-\frac{9}{4}\bar{\delta}^2.
\end{eqnarray}

In slow-roll inflation, the spectral index  and its running are given by
\begin{eqnarray}\label{x15}
 n_s-1=2\bar\eta_{sr}-6\bar\epsilon_{sr}, \quad \alpha_s=16\bar\epsilon_{sr}\bar\eta_{sr}-24\bar\epsilon_{sr}^2-2\bar\xi^2_{sr},
\end{eqnarray}
where $\bar\xi^2_{sr}\cong \frac{V_{\varphi}V_{\varphi\varphi\varphi}}{V^2}$ is a second order slow roll parameter .

%\end{eqnarray}
Moreover, in inflationary cosmology, the scalar/tensor ratio is related with the slow-roll parameter $\bar\epsilon_{sr}$, in the way
\begin{eqnarray}\label{x18}
r=16\bar\epsilon_{sr},
\end{eqnarray}
what does not happen in the matter bounce scenario, because there the tensor/scalar ratio depends on the whole background dynamics, and not solely on those corresponding to 
quasi-matter domination.

\subsection{Power law expansion}

As an example,
we will choose the following potential \cite{Lucchin-Matarrese}
\begin{eqnarray}\label{xpotential}
V(\varphi)=V_0 e^{-\sqrt{3(1+\omega)}|\varphi|},
\end{eqnarray}
which leads to the power law expansion
\begin{eqnarray}a\propto t^{\frac{2}{3(1+\omega)}}.\end{eqnarray}
An easy calculation yields, for the matter bounce scenario,
\begin{eqnarray}\label{x13}
 n_s-1=12\omega.
\end{eqnarray}
On the contrary, in the case of slow-roll inflation, for the same potential (\ref{xpotential}), one gets
\begin{eqnarray}
 %\bar\epsilon_{s-r}=\frac{3}{2}(1+\omega),\quad\bar\eta_{s-r}={3}(1+\omega)\Longrightarrow
 n_s-1=-{3}(1+\omega)\qquad \mbox{and}\qquad r=24(1+\omega).
\end{eqnarray}

For any of these theories, matter bounce scenario and inflation, to be viable they have to match more and more accurate astronomical data. Focussing, in particular, on PLANCK data, the 
resulting spectral index is given by $n_s=0.9603\pm 0.0073$, what specifically means that:
\begin{enumerate}
 \item  In the matter bounce scenario, in order for the potential (\ref{xpotential}) to match with observations, one needs to  choose $\omega=-0.0033\pm 0.0006$.
 \item In power law  inflation, the  potential (\ref{xpotential}) turns out to be in agreement with the observational value of the spectral index provided $\omega=-0.9867\pm 0.0024$.
  Moreover,  since the tensor/scalar ratio is given by
$r=24(1+\omega)$, for this potential to fit well with PLANCK data one has to impose $\omega\leq -0.9954$, which is not compatible with the previous number $\omega=-0.9867\pm 0.0024$.
On the other hand, to match the ratio of tensor to scalar perturbations with the  BICEP2 data, one has to choose $\omega\in [-0.9937,-09887]$ what, together with the condition
$\omega=-0.9867\pm 0.0024$, restricts the value of the parameter $\omega$ to be $\omega=-0.9890^{+0.0001}_{-0.0003}$.
\end{enumerate}
This calculation clearly shows that, in order to match with current observational data, the parameter $\omega$ which appears in both theories, must be conveniently tuned.

Finally, the power law expansion given by the potential (\ref{xpotential})
has no running, what is in contradiction with the very last PLANCK data \cite{Ade}, which provides the following experimental value $\alpha_s=-0.0134\pm 0.009$.
For this reason, some other models must be alternatively considered.

\section{Quasi-matter domination potentials obtained from the equation of state}

We now continue, once again  with the parametrization of the scale factor in the contracting phase given by $a(N)=a_fe^N$, where $a_f$ is the value of the scale factor at the end of the
quasi-matter domination
period. We will assume, as in inflation \cite{mukhanov},  an equation of state (EoS) of the form $\frac{P}{\rho}=\frac{\beta}{(N+1)^{\alpha}}$ where
$\alpha>0$ and $\beta<0$ (the fluid has negative pressure) are
both  of order 1. This particular dependence between  $\frac{P}{\rho}$ and the number of e-folds, will allows us to obtain, in a simple way, potentials 
that lead to a quasi-matter domination. 
Effectively,
the conservation equation reads
\begin{eqnarray}
 \frac{d\ln\rho}{dN}=-3\left(1+\frac{P}{\rho} \right)=-3-\frac{3\beta}{(N+1)^{\alpha}},
\end{eqnarray}
and the solution of this equation is given by
\begin{eqnarray}\label{y3}
 \rho(N)=\left\{\begin{array}{cc}
 \rho_f e^{-3N}(N+1)^{-3\beta},&  \alpha=1,\\
 \rho_0 e^{-3N}e^{\frac{3\beta}{(\alpha-1)(N+1)^{\alpha-1}}},&  \alpha\not=1.
 \end{array}\right.
\end{eqnarray}
On the other hand, in the contracting phase
\begin{eqnarray}
 \frac{d\varphi}{d N}=\frac{\dot{\varphi}}{H}=-\sqrt{3}\sqrt{1+\frac{P}{\rho} }=-\sqrt{3}\sqrt{1+\frac{\beta}{(N+1)^{\alpha}}},
\end{eqnarray}
where we have used that $\dot{\varphi}^2=\rho+P$. This equation could be explicitly integrated for $\alpha=1,2$,
for example, when $\alpha=1$, one has
\begin{eqnarray}\label{alpha}
\varphi(N)=-\sqrt{3\left(1+\frac{\beta}{(N+1)}\right)}(N+1)+\frac{\sqrt{3}\beta}{2}\ln\left(\frac{\sqrt{1+\frac{\beta}{(N+1)}}-1}{\sqrt{1+\frac{\beta}{(N+1)}}+1   }\right).
\end{eqnarray}
But here we will do the approximation
$\sqrt{1+\frac{\beta}{(N+1)^{\alpha}}}=1+\frac{\beta}{2(N+1)^{\alpha}}$, for large values of $N$. Then, one gets
\begin{eqnarray}\label{y2}
 \varphi(N)\cong \left\{\begin{array}{cc}
  -\sqrt{3}\left(N +\ln (N+1)^{\frac{\beta}{2}}\right),& \alpha=1,\\
    -\sqrt{3}\left(N- \frac{\beta}{2(\alpha-1)(N+1)^{\alpha-1}}\right), & \alpha\not=1
    \end{array}\right.
 %\cong -\sqrt{3}N.
\end{eqnarray}

Finally, introducing the approximation of quasi-matter domination, $\rho(N)\cong 2V(N)$, we obtain that, for $N\geq 1\Longleftrightarrow \varphi\rightarrow -\infty$,
\begin{eqnarray}\label{potential}
 V(\varphi)\cong \left\{\begin{array}{cc}
 V_0e^{\sqrt{3}\varphi}(N(\varphi)+1)^{-\frac{3\beta}{2}},& \alpha=1\\
 V_0e^{\sqrt{3}\varphi} e^{\frac{3\beta}{2(\alpha-1)(N(\varphi)+1)^{\alpha-1}}},&\alpha\not=1,
 \end{array}\right.
\end{eqnarray}
where $N(\varphi)$ is got by solving for $N$ in (\ref{y2}).

\subsection{Viability of the models}

The spectral index and the running parameter for the EoS $\frac{P}{\rho}=\frac{\beta}{(N+1)^{\alpha}}$, and thus for potentials of the form (\ref{potential}),
can be easily obtained from Eq.~(\ref{espectral_index}) by using the relation $\frac{d}{Hdt}=\frac{d}{dN}$, what yields
%for large values of $N$
\begin{eqnarray}\label{espectral2}
 n_s-1
 %=
 %4\left(3\frac{P}{\rho}-\frac{1}{2}\frac{d}{dN}\ln\left(1+\frac{P}{\rho} \right)\right)
 %\cong
 =\frac{12\beta}{(N+1)^{\alpha}}, \quad \alpha_s\cong \frac{24\alpha\beta}{(N+1)^{\alpha+1}}.
\end{eqnarray}
Note that, in the matter bounce scenario, the ratio of tensor to scalar perturbations  is not related with the quasi-matter domination parameters and has to be calculated
using Eq.~(\ref{ratio}). This calculation can be carried out numerically for  the solution of the conservation equation
\begin{eqnarray}\label{conservation}
 \ddot{\varphi}+3H(\varphi)\dot{\varphi}+V_{\varphi}=0,
\end{eqnarray}
corresponding to a Universe that takes $N$ e-folds to leave the quasi-matter domination epoch,
%In fact, it has to be done for  the solution of (\ref{conservation}) that, at early times depict a quasi-matter dominated Universe in the contracting phase,
i.e., for the solution which satisfies the initial conditions:
\begin{eqnarray}
 \varphi_i=\varphi(N),\qquad \dot{\varphi}_i=H\frac{d\varphi}{d N}=\sqrt{\rho(N)}\sqrt{1+\frac{\beta}{(N+1)^{\alpha}}},
\end{eqnarray}
where $\varphi(N)$ and $\rho(N)$ are given by (\ref{y2}) and   (\ref{y3}), respectively.

%These theoretical results must be then compared with the astronomical ones (e.g., from PLANCK or BICEP2) to test the viability of the models.

However, it is important to realize that
the constrain of the tensor/scalar ratio provided by WMAP and PLANCK projects (${r}\leq 0.11$) is obtained indirectly assuming the  {\it consistency} slow roll relation
${r}=16\bar{\epsilon}_{sr}$
%(where $\bar{\epsilon}=-\frac{\dot{H}}{H^2}\cong \frac{1}{2}\left(\frac{V_{\varphi}}{V} \right)^2$ is the main slow roll parameter)
 \cite{Peiris}, because
gravitational waves
are not detected by those projects. This means that the slow roll inflationary models must satisfy this constrain, but
not the bouncing ones, where there is not any consistency relation.
This point is very important because some very complicated mechanisms are sometimes implemented in the MBS  in order to enhance the
power spectrum of scalar perturbations to achieve the observational bound provided by PLANCK \cite{Cai1}. Moreover,  numerical calculation have been
performed for holonomy corrected and teleparallel Loop Quantum Cosmology \cite{Haro3}, and those theoretical values of the tensor/scalar ratio have been compared with the
corresponding observational values
provided by PLANCK and BICEP2 projects.
In fact, in matter bounce scenario, to check  if the models provide a viable value of the
tensor/scalar ratio, first of all gravitational waves must be clearly detected in order to determine the observed value of this ratio. We
hope that more accurate unified PLANCK-BICEP2 data (the B2P collaboration), which is going to be issued soon, may
address this point. In contrast,  the spectral index of scalar perturbations
and its running could be calculated independently of the theory \cite{verde}, which means that in order to check bouncing models, while in the absence of evidence  of gravitational waves, 
one
has to work in the space $(n_s,\alpha_s)$.

\vspace{3mm}

\subsubsection{Example 1.} As a first example,
we can compare our results relative to the matter bounce scenario with those for chaotic inflation given by the  potential $V(\varphi)=\lambda \varphi^{2n}$. In this case, one has \cite{haro-amoros}
\begin{eqnarray}\label{y1}
 n_s-1=-\frac{2(n+1)}{2N+n},\qquad r=\frac{16n}{2N+n}.
\end{eqnarray}
We can see that,  for the same number of e-folds, one obtain the same spectral index in both the matter bounce scenario and chaotic inflation, after choosing
in the matter bounce scenario $\alpha=1$ and $\beta=-\frac{1}{4}$, and a quartic potential for inflation. That is, for these parameters $\alpha=1$, $\beta=-\frac{1}{4}$ and $n=2$,
for modes which leave the Hubble radius {}{about} a number $N$ of e-folds before the end of the corresponding period (quasi-matter domination in bouncing cosmologies and the
slow-roll phase in inflation), one
obtains {\it the same} spectral index.

From (\ref{y1}) we can see that in order
to achieve the observed value of the spectral index one has to choose $N\in [62.829,91.592]$.
On the other hand, in slow roll inflation one also has the constrain $r=\frac{16}{N+1}$, what compared with the PLANCK  constrain $r\leq 0.11$ implies
$N\geq 144.454$. But this means that the chaotic quartic potential is  {}{ruled out} 
%\sout{disfavored} 
by PLANCK data.
However, if one considers the BICEP2 data $r=0.20^{+0.07}_{-0.05}$, one obtains that for $N\in [58.259,105.666]$ what means that  the quartic potential fits well with BICEP2
data for $N\in [62.829,91.592]$.

For our bouncing model, with the aim to obtain the theoretical value of the  tensor/scalar ratio one has to use the formula (\ref{ratio}). Thus, one needs to calculate this quantity for 
the solution of the conservation equation with initial conditions
\begin{eqnarray}
 \varphi_i= \varphi(N), \quad \dot{\varphi}_i=H\frac{d\varphi}{d N}=\sqrt{\rho(N)}\sqrt{1-\frac{1}{4(N+1)}},
\end{eqnarray}
where $\varphi(N)$ is given by (\ref{alpha}),  $\rho(N)$  by (\ref{y3}) and  $N\in [62.829,91.592]$.

\vspace{3mm}

\subsubsection{Example 2.} As a seconde example we will deal with $R^2$ gravity, where \cite{Odintsov}
\begin{eqnarray}
 n_s-1=-\frac{2}{N},\quad r=\frac{12}{N^2}.
\end{eqnarray}
The same spectral index could be obtained from (\ref{espectral2}) choosing $\alpha=1$, $\beta=-\frac{1}{6}$ and considering $N-1$ e-folds, instead of $N$.
In this case the correct power spectrum is obtained by choosing $N\in [42.553, 61.728]$.

In inflationary cosmology the model matches correctly with PLANCK data, because the constrain $r\leq 0.11$ is equivalent to
$N\geq 10.44$. However, the model is incompatible with the BICEP2 data, because it implies $N\in [6.66,8.94]$.

\subsection{Compatibility between the spectral index and the running parameter}

Here we will compare the compatibility of the spectral index and the running parameter in both the inflation and matter bounce scenarios. In the slow-roll regime, both the spectral
index and the running parameter for a perfect fluid can be easily calculated from the equations
\begin{eqnarray}\label{c}
 n_s-1=-3\left(1+\frac{P}{\rho} \right)+\frac{d}{dN}\ln\left(1+\frac{P}{\rho} \right), \quad \alpha_s=\frac{\dot{n}_s}{H}=-\frac{d n_s}{dN},
\end{eqnarray}
and if one considers a fluid satisfying the condition
\begin{eqnarray}\label{condition}\left|\frac{d}{dN}\ln(\rho+P)\right|\ll \left|1+\frac{P}{\rho}\right|,
\end{eqnarray}
one gets  \cite{odintsov}
\begin{eqnarray}
 n_s-1\cong -6\left(1+\frac{P}{\rho} \right), \quad \alpha_s\cong -18\left(1+\frac{P}{\rho} \right)^2.
\end{eqnarray}

From these equations one obtains the relation $\alpha_s=-\frac{1}{2}\left({1-n_s}\right)^2$. Now, inserting the observed value for the spectral index, $n_s=0.9603\pm 0.0073$, this yields
%$\alpha_s=-4\times 10^{-4}\pm 1.4\times 10^{-4}$
$\alpha_s\in (-5.2\times 10^{-4}, -1.1\times 10^{-3})$
which is in clear contradiction with the observed value $\alpha_s=-0.0134\pm 0.009$. As a consequence, inflation corresponding to this kind of  perfect
fluid is less favored by the current observational data.

For a fluid with an EoS $1+\frac{P}{\rho}=\frac{\beta}{(N+1)^{\alpha}}$, with both $\alpha$ and $\beta$ positive and of order $1$ \cite{mukhanov}, which does not satisfy the condition (\ref{condition}), one has
\begin{eqnarray}
 \alpha_s=\left\{\begin{array}{cc}
  -\frac{1}{\alpha}(n_s-1)^2& \alpha>1\\
     -\frac{1}{3\beta+1}(n_s-1)^2& \alpha=1\\
      \frac{\alpha}{N+1}(n_s-1)& \alpha<1.
                \end{array}
\right.
\end{eqnarray}
%what means that $|\alpha|\leq (n_s-1)^2=(0,0397\pm 0.0073)^2\ll | -0.0134\pm 0.009|$, rouling out this kind of fluids.
Then,
\begin{enumerate}\item
for $\alpha\geq 1$ one has $$|\alpha_s|\leq (n_s-1)^2=(0,0397\pm 0.0073)^2< | -0.0134\pm 0.009|,$$ ruling out, at $1\sigma$ confidence level, this kind of models.
\item for $\alpha< 1$, to match the theoretical values with the experimental ones, the parameters must satisfy
$$2\alpha\leq N+1\leq 11\alpha\quad\mbox{and}\quad -\frac{3\beta}{(N+1)^{\alpha}}=-0.0397\pm 0.0073.
$$

However, since $\alpha<1$ the number of e-folds before the end of inflation satisfy $N+1<11$, which is incompatible with the minimum number of e-folds (for the most general
models $N\geq 50$ \cite{Ade})
to solve both the horizon and flatness problems in General Relativity.

%and thus, the ratio of tensor to scalar perturbations is bounded by 
%\begin{eqnarray}
% r=\frac{24\beta}{(N+1)^{\alpha}}\geq \frac{24\beta}{11},
%\end{eqnarray}
%which is incompatible with the PLANCK bound $r\leq 0.11$, because both bounds are only satisfied when $\beta \leq 0.05$ what is incompatible with the assumption $\beta\sim {\mathcal O}(1)$.

%However, those bounds are never obtained for $N\geq 50$ (the minimum number of e-folds needed to solve both the horizon and flatness problems in General Relativity). In fact, for $N=50$ one has
%$\alpha\geq 4,63$, and consequently $-\frac{3\beta}{(N+1)^{\alpha}}\leq 4.2\times 10^{-7}\beta$. Then, since $\beta$ is of order $1$, the value $ -0.0397\pm 0.0073$
%is never reached, what again means that this kind of models are less favored by present observational data.

%This bounds are easily satisfied. For example, choosing $N=60$, $\alpha=\frac{3}{4}$ and $\beta=\frac{1}{4}$ one obtains
%\begin{eqnarray}
% n_s=0,9652
%\end{eqnarray}

\end{enumerate}

The problem with slow-roll inflation is that, in general, the spectral index is of the order $N^{-1}$, while the running parameter is of order $N^{-2}$ and, consequently,   one has
$\alpha_s\sim (1-n_s)^2$, which in most cases is incompatible with PLANCK data, because the observed value of the running is not small enough \cite{running}.
Moreover,
the constrain of the tensor/scalar ratio provided by WMAP and PLANCK projects (${r}\leq 0.11$) is obtained indirectly assuming the  {\it consistency} slow roll relation
${r}=16\bar{\epsilon}_{sr}$ \cite{Peiris}, because
gravitational waves
are not detected by those projects. This means that the slow roll inflationary models must satisfy this constrain, but
not the bouncing ones, where there is not any consistency relation.
And it is the combination of the three data $(n_s,\alpha_s,{r})$ what 
rules out, at $1\sigma$ confidence level for the running,  all the standard slow-roll inflationary models.

{{Effectively,  for instance, we consider the $\Lambda$CDM$+r+\alpha_s$ model from PLANCK combined with WP and BAO data, which gives the following results
$n_s=0.9607\pm 0.0063$, $r\leq 0.25$ at $95$\% C.L. and $\alpha_s=-0.021^{+0.012}_{-0.010}$ (see table 5  of $\cite{Ade}$).
In slow roll inflation, a simple calculation leads to the relation
\begin{eqnarray}
\alpha_s=\frac{1}{2}(n_s-1)r+\frac{3}{32}r^2-2\bar{\xi}^2_{sr}.
\end{eqnarray}

And thus, considering $n_s$ at 2$\sigma$ confidence level and taking the conservative bound $r\leq 0.32$ (see Fig.~4 of $\cite{Ade}$),  the minimum of the function
$ \frac{1}{2}(n_s-1)r+\frac{3}{32}r^2$ 
is bigger than $-0.0018$, what provides the bound
\begin{eqnarray}\label{X}
\alpha_s\geq -0.0018-2\bar\xi^2_{sr};
\end{eqnarray}
what means that potentials such as $V(\varphi)=V_0\left(1-\frac{\varphi^2}{\mu^2}+\dots\right)$ 
(hilltop), $V(\varphi)=V_0\left(1-\frac{\varphi^2}{\mu^2}\right)^2$ (plateau) \cite{olive} or
$V(\varphi)=V_0\left(1+\cos\left(\frac{\varphi}{\mu}\right)\right)$ (natural) \cite{freese}, 
when one considers values of the running at 1$\sigma$ confidence level ($\alpha_s=-0.021^{+0.012}_{-0.010}\Longleftrightarrow -0.031\leq \alpha_s\leq -0.009$),
are disfavored
by PLANCK data because for all of them $\bar\xi^2_{sr}\leq 0$.

Dealing with the monomial  potential $V(\varphi)=V_0\varphi^{p}$, one obtains
\begin{eqnarray}\label{zzw}
 n_s-1=-\frac{p(p+2)}{\varphi^2},\quad \alpha_s=-\frac{2p^2(p+2)}{\varphi^4},
\end{eqnarray}
what means that $p$ must be positive in order to have an spectral index with a red tilt and a negative running. As a first consequence, inverse power law potentials \cite{barrow} are disfavored.

For $p=1,2$ one has $\bar\xi^2_{sr}=0$, and thus, one can apply the bound (\ref{X}) to disfavor these models.
In general, since 
for this monomial potential one has
$\bar\xi^2_{sr}=\frac{p-2}{p-1}\bar\eta^2_{sr}$, one can obtain 
the following exact formula
\begin{eqnarray}
\alpha_s=\frac{p+2}{8(p-1)}(n_s-1)r+\frac{3(p+2)}{128(p-1)}r^2-\frac{p-2}{2(p-1)}(n_s-1)^2.
\end{eqnarray}

And for $p\geq 3$, using that $\frac{p-2}{p-1}\leq 2$ and the fact that $\frac{p+2}{p-1}$ increases as a function of $p$, one gets the bound
\begin{eqnarray}
\alpha_s\geq\frac{5}{16}(n_s-1)r-(n_s-1)^2\geq -0.0084,
\end{eqnarray}
which is incompatible with the running provided by PLANCK at $1\sigma$ confidence level.

Finally, for a general hilltop potential $V(\varphi)=V_0\left(1-\frac{\varphi^p}{\mu^p}+\dots\right)$ with $p\geq 3$ \cite{albrecht}, one also has the relation
$\bar\xi^2_{sr}=\frac{p-2}{p-1}\bar\eta^2_{sr}$, and thus, one can apply the same reasoning as in the previous case for  monomial potentials.
}}

A way to solve this problem
is to break the slow-roll approximation  for a short while, as due, for example, to the
inclusion of a quickly oscillating term in the potential. In this case the theoretical value of the running parameter gets larger and could match well with experimental data \cite{Cai}.

%\vspace{1cm}

On the other hand, in the matter bounce scenario, when dealing with a perfect fluid with EoS $\frac{P}{\rho}=\frac{\beta}{(N+1)^{\alpha}}$, one obtains from (\ref{espectral2})
the following relation
\begin{eqnarray}\alpha_s=\frac{2\alpha}{N+1}(n_s-1)\end{eqnarray}
which is perfectly compatible with the experimental data.
In fact, for instance, if one takes $\alpha=2$ and $N=12$, 
(note that in bouncing cosmologies a large number of e-folds is {\it not} required, because the horizon problem does not exist, since
at the bounce all  parts of the Universe are already in causal contact,
and also the flatness problem gets improved \cite{brandenberger1}),
one obtains, for $n_s=0.9603\pm 0.0073$, the following value for the running parameter: $\alpha_s=0.0122\pm 0.0022$, which is
compatible with the PLANCK data. Effectively, for these values of $\alpha$ and $N$ one gets $n_s-1=\frac{12}{13^2}\beta\cong 0.071\beta$, which is indeed compatible with its
observed value, by choosing $\beta\cong-\frac{1}{2}$.

\section{Conclusions}

In this paper we have introduced, at the early times in the contracting phase of bouncing cosmologies,  a quasi-matter domination regime controlled by some convenient small parameters which we have defined here. This has allowed us to
obtain theoretical  values of the spectral index and of the running parameter which are in perfect agreement with the most recent and accurate observational data from the PLANCK satellite.

We have shown in detail, and with the help of several simple examples, the viability of our bouncing models for isotropic fluids with an equation of state which depends on the number of 
e-folds occurring before the end of the quasi-matter domination epoch. We have also demonstrated that, in contrast to these results, slow-roll inflationary models are generically less favored by the most recent PLANCK observational data due, in particular, to the rather small  value  of the running parameter predicted by all these slow-roll theories.

We expect that more precise unified PLANCK-BICEP2 data (the B2P collaboration), which are going to be issued soon, may even fit better the 
bouncing cosmologies under consideration here.

\vspace{1cm}
{\bf Acknowledgments.}
This investigation has been supported in part by MINECO (Spain), projects MTM2011-27739-C04-01, FIS2010-15640 and FIS2013-44881, and by the CPAN Consolider Ingenio Project.

\end{document}